\documentclass[prl,aps,twocolumn,showpacs]{revtex4-1}
\usepackage{graphicx}
\usepackage{amsmath}
\usepackage{bm}
\usepackage{amstext}
\usepackage{amsxtra}

\usepackage{times}

\begin{document}

\newcommand{\To}{T_c^0}
\newcommand{\kB}{k_{\rm B}}
\newcommand{\dT}{\Delta T_c}
\newcommand{\lo}{\lambda_0}
\newcommand{\cs}{$\clubsuit$}
\newcommand{\thold}{t_{\rm hold}}
\newcommand{\Nmf}{N_c^{\rm MF}}
\newcommand{\Tmf}{T_c^{\rm MF}}
\newcommand{\mumf}{\mu_c^{\rm MF}}
\newcommand{\nmf}{n_c^{\rm MF}}

\title{Condensed Fraction of an Atomic Bose Gas Induced by Critical Correlations}

\author{Robert P. Smith$^1$, Naaman Tammuz$^1$, Robert L. D. Campbell$^1$,  Markus Holzmann$^{2,3}$,
and Zoran Hadzibabic$^1$}
\affiliation{$^1$Cavendish Laboratory, University of Cambridge, J.~J.~Thomson Avenue, Cambridge CB3~0HE, United Kingdom \\
$^2$LPTMC, UMR 7600 of CNRS, Universit\'e P. et M. Curie, 75752 Paris, France \\
$^3$LPMMC, UMR 5493 of CNRS, Universit\'e J. Fourier, 38042 Grenoble, France}

\begin{abstract}
We study the condensed fraction of a harmonically-trapped atomic Bose gas at the critical point predicted by mean-field (MF) theory. The non-zero condensed fraction $f_0$ is induced by critical correlations which increase the transition temperature $T_c$ above $\Tmf$.
Unlike the $T_c$ shift in a trapped gas, $f_0$ is sensitive only to the critical behaviour in the quasi-uniform part of the cloud near the trap centre.
To leading order in the interaction parameter $a/\lo$, where $a$ is the s-wave scattering length and $\lo$ the thermal wavelength, we expect a universal scaling $f_0 \propto (a/\lo)^4$. We experimentally verify this scaling using a Feshbach resonance to tune $a/\lo$. Further, using the local density approximation, we compare our measurements with the universal result obtained from Monte-Carlo simulations for a uniform system, and find excellent quantitative agreement.
\end{abstract}

\date{\today}

\pacs{03.75.Hh, 67.85.-d}

%03.75.Hh	Static properties of condensates; thermodynamical, statistical, and structural properties
%67.85.-d 	Ultracold gases, trapped gases
%03.75.Kk 	Dynamic properties of condensates; collective and hydrodynamic excitations, superfluid flow
%Optical angular momentum (quantum optics), 42.50.Tx
%47.37.+q 	Hydrodynamic aspects of superfluidity; quantum fluids
%67.85.De 	Dynamic properties of condensates; excitations, and superfluid flow
%37.10.Vz 	Mechanical effects of light on atoms, molecules, and ions

\maketitle

Some of the most interesting fundamental problems of many-body physics involve strong inter-particle correlations, and cannot be addressed by mean-field (MF) theories.
Harmonically trapped ultracold atomic gases are promising candidates for highly controllable ``quantum simulation" of such intricate many-body scenarios \cite{Bloch:2008}. However, for testing the existing theories of spatially uniform systems, it is often important to experimentally extract information on local properties of a non-uniform trapped gas (see e.g. \cite{Ho:2010a, Nascimbene:2010, Yefsah:2011}).

\begin{figure} [b]
\includegraphics[width=\columnwidth]{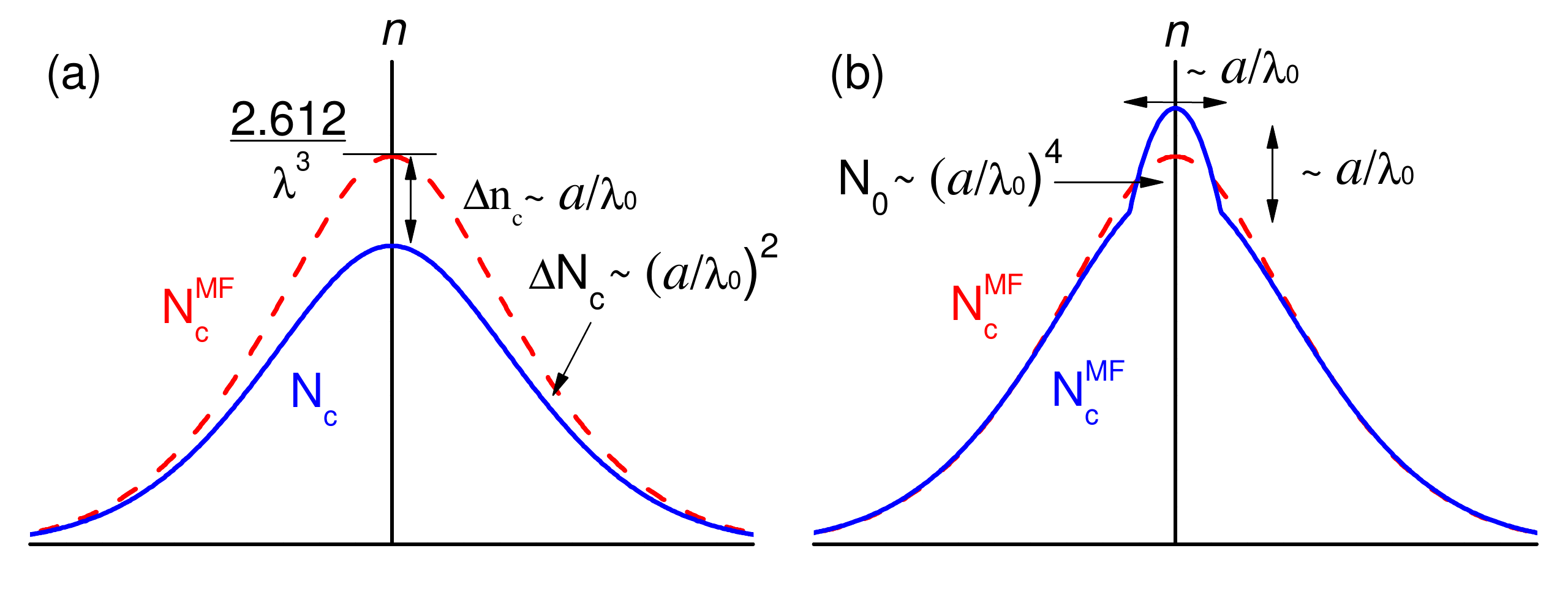}
\caption{(color online) Beyond-mean-field effects near the critical point in a harmonically trapped Bose gas. (a) For a fixed temperature, the density distribution at the critical point $N=N_c < \Nmf$ (solid blue line) is compared with the mean-field prediction (dashed red line). In the trap centre we expect $\nmf - n_c  \propto a/\lo$, characteristic of the critical behaviour in a uniform system. However the experimentally measured $N_c$ shift, $\Nmf - N_c \propto (a/\lo)^2$, is dominated by the density shift {\it outside} the central critical region, and is not directly related to the $n_c$ shift.
(b) If $N$ is increased to $\Nmf > N_c$, a small condensate induced by critical correlations forms within the critical region of size $\propto a/\lo$. The condensed atom number  $N_0 \propto (a/\lo)^4$ directly relates to the critical density shift $\Delta n_c \propto a/\lo$.}
\label{fig:Cartoon}
\end{figure}

The effect of interactions on Bose-Einstein condensation of a dilute gas is a classic example of a difficult beyond-MF problem, which has challenged theorists for decades \cite{Lee:1957TcShift, Lee:1958TcShift2, Bijlsma:1996, Baym:1999, Holzmann:1999, Reppy:2000, Arnold:2001, Kashurnikov:2001, Holzmann:2001, Baym:2001, Kleinert:2003, Andersen:2004RMP, Holzmann:2004}.
It is also an example of a situation where harmonic confinement both quantitatively and qualitatively alters the physics \cite{Giorgini:1996, Ensher:1996,  Houbiers:1997, Holzmann:1999b, Arnold:2001b, Gerbier:2004, Davis:2006, Zobay:2009, Meppelink:2010, Smith:2011}.
For a uniform gas the interaction shift of the critical temperature $T_c$ cannot be calculated to any order in the interaction strength using perturbation theory, owing to strong correlations that develop near the critical point.
On the other hand, non-uniformity of a trapped atomic gas results in a significant MF shift of $T_c$ \cite{Giorgini:1996}. More importantly, it diminishes the more interesting beyond-MF effects, in essence because near $T_c$ only a small fraction of the cloud is actually in the critical regime (see Fig.~\ref{fig:Cartoon}). Only recently have the beyond-MF effects on condensation of an atomic gas become experimentally accessible  \cite{Smith:2011}.
Many questions remain open since beyond-MF effects in a uniform and a trapped gas have different dependence on the strength of interactions, and quantitative connections between the two are highly non-trivial.

In this Letter, we study the condensed fraction ($f_0$) of an atomic Bose gas at the critical point predicted by MF theory.
By definition $f_0$ vanishes within MF theory, and directly measures the effect of critical correlations which shift $T_c$ above $\Tmf$.
Moreover, while the $T_c$ shift itself strongly depends on the global properties of a non-uniform gas, $f_0$ measurements directly probe the quasi-uniform critical region near the centre of the trap.
To leading order in the strength of interactions we predict a universal scaling $f_0 \propto (a/\lo)^4$, where $a>0$ is the s-wave scattering length and $\lo$ the thermal wavelength at the ideal gas critical temperature $\To$. Using a Feshbach resonance in a $^{39}$K gas to tune $a/\lo$, and accurately measuring condensed fractions in the range $0.1 - 1\%$, we experimentally verify this prediction. Further, we directly relate our measurements to the universal critical behaviour seen in the classical-field Monte-Carlo simulations of a {\it uniform} system \cite{Prokofev:2004}, and find excellent quantitative agreement.

In Fig.~\ref{fig:Cartoon}(a) we illustrate the difference between the beyond-MF shifts of the critical point in a uniform and a trapped system, and in
Fig.~\ref{fig:Cartoon}(b) the expected scaling of the condensed fraction at the MF critical point. For visual clarity, here we fix the temperature of the gas and consider the interaction shift of the critical density $n_c$ (in the centre of the trap) and the critical atom number $N_c$.
Surprisingly, the beyond-MF shifts of $n_c$ and $N_c$ are not directly related to each other.
The quadratic beyond-MF $N_c$ shift is of direct relevance to the experimentally pertinent case of a trapped gas, but from the point of view of the theory of critical behaviour the linear $n_c$ shift is actually more interesting. Here we will show how to use a trapped atomic cloud to experimentally obtain information about the critical behaviour in a uniform system.

We first outline some general scaling arguments, then present our experimental results, and finally return to a quantitative comparison of our measurements with the theory based on the classical-field Monte-Carlo simulations of Ref. \cite{Prokofev:2004} for a uniform system.

In a uniform system, ideal-gas condensation occurs at a chemical potential $\mu_c^0 =0$, and a critical phase space density $n\lambda^3 = \zeta(3/2) \approx 2.612$, where $\zeta$ is the Riemann function. In an interacting gas there is no $T_c$ shift at MF level, i.e. $\Tmf = \To$. To leading order in $a/\lo \ll 1$ the expected beyond-MF $T_c$ shift is given by \cite{Bijlsma:1996, Baym:1999, Holzmann:1999, Reppy:2000, Arnold:2001, Kashurnikov:2001, Holzmann:2001, Baym:2001, Kleinert:2003, Andersen:2004RMP, Holzmann:2004}
\begin{equation}
\frac{\dT}{\To} \approx  c \, \frac{a}{\lo} \; ,
\label{eq:UniTc}
\end{equation}
where $\dT = T_c - \To$, and $c \approx 1.8$ \cite{Arnold:2001, Kashurnikov:2001}.
Equivalently, the $n_c$ shift at constant $T$ is $\Delta n_c/n_c^0 \approx - (3/2) \dT /\To$.

An important point is that, at both MF and beyond-MF level, the interactions differently affect $T_c$ (or equivalently $n_c$) and the critical chemical potential $\mu_c$. The simple MF shift $\beta \mumf = 4 \, \zeta(3/2) a/\lo$, where $\beta = 1/\kB T$, has no effect on condensation, and to lowest beyond-MF order \cite{B2}:
\begin{equation}
\beta \mu_c \approx \beta \mumf + B_2 \left( \frac{a}{\lo} \right)^2 \; .
\label{eq:Muc}
\end{equation}
The qualitative difference between
Eqs. (\ref{eq:UniTc}) and (\ref{eq:Muc}) highlights the fact that the problem of the $T_c$ shift is non-perturbative and near criticality the equation of state does not have a regular expansion in $\mu$ (otherwise one would get $\Delta n_c \propto \mu_c^{MF} - \mu_c$).

In a harmonically trapped gas $T_c$ is defined for a given atom number $N$, rather than for a given density $n$. For an ideal gas $\kB \To = \hbar \omega \, \left[ N/\zeta(3) \right]^{1/3}$, where $\zeta(3) \approx 1.202$.
Within the local density approximation (LDA) one expects the uniform-system results for $n_c$ and $\mu_c$ to apply in the centre of the trap, $\mathbf{r} = 0$. Elsewhere in the trap  the local chemical potential is $\mu( \mathbf{r} ) = \mu(0) - m\omega^2 r^2 / 2$, where $m$ is the atom mass and $\omega$ the trapping frequency.
The result for the $T_c$ shift however does not carry over so easily to the trapped case; the basic reason for this is that at $T_c$ only a small fraction of the non-uniform cloud is actually in the critical regime. The size of the central critical region is $r_c \sim (a/\lo) R_T$, where $R_T = \sqrt {\kB T / m\omega^2}$ is the thermal radius \cite{Arnold:2001b}. Combining this with $\Delta n_c \sim a/\lo$ implied by Eq.~(\ref{eq:UniTc}), we obtain a very small beyond-MF shift of the critical number of atoms within the critical region, of the order $(a/\lo)^4$. However the interaction shift of $\mu_c$ affects the density everywhere in the trap.
One famous consequence of this is the negative MF shift of $T_c$ in a harmonically trapped gas \cite{Giorgini:1996}: while
$\nmf = n_c^0$, repulsive interactions broaden the density distribution so that $\Nmf > N_c^0$.

More generally, the experimentally observed $T_c$ shift in a trapped gas \cite{Smith:2011} qualitatively mirrors Eq.~(\ref{eq:Muc}):
\begin{equation}
\frac{\dT}{\To} \approx b_1 \, \frac{a}{\lo} + b_2 \left( \frac{a}{\lo} \right)^2\, .
\label{eq:TrapTc}
\end{equation}
Here $b_1 \approx -3.426$ is an analytical, strictly MF result \cite{Giorgini:1996}, and $b_2 = 46 \pm 5$ was measured in \cite{Smith:2011}. The quadratic $T_c$ shift depends on the beyond-MF correlations, but does not directly correspond to the lowest-order (beyond-MF) $T_c$ shift in a uniform system [Eq.~(\ref{eq:UniTc})], which should enter the in-trap $T_c$ only at the $(a/\lo)^4$ level. We can qualitatively understand the similarity of Eqs.~(\ref{eq:Muc}) and (\ref{eq:TrapTc}) by noting that: (i) Away from the critical point the equation of state is regular in $\mu$ and the local density shift is simply proportional to $\mu_c$, at both MF and beyond-MF level \cite{details}, and (ii) The contribution to $N_c$ from the non-critical region outweighs the contribution from within the critical region by a large factor $\sim (\lo/a)^3$.

To summarize this analysis: On the one hand we expect $\nmf - n_c  \propto a/\lo$,
characteristic of the critical behaviour in a uniform system. On the other hand $\Nmf - N_c \propto (a/\lo)^2$ is dominated by the effect of the $\mu_c$ shift on the density \emph{outside} the critical region. The latter result was observed in \cite{Smith:2011}; the former cannot be experimentally verified without a direct probe of the local density in a 3D cloud.

By studying the condensed fraction $f_0$ at the MF-predicted critical point we overcome the problem of the absence of the local density probe, and gain more direct insight into the critical behaviour in the centre of the trap. Simply put, instead of asking how $N_c$ is reduced with respect to $\Nmf$ by critical correlations, we ask how many atoms pile up in the condensate if (at constant $T$) we increase the total atom number to $\Nmf > N_c$. Experimentally, the obvious advantage is that the condensed and thermal component can be clearly distinguished in standard time-of-flight (TOF) expansion, thus allowing us to use a ``global" measurement technique to access the local behaviour of the gas within the critical region. Theoretically, the analogous quantity for a uniform gas, $n_0/n$ (where $n_0$ is the condensate density), was first considered by Holzmann and Baym \cite{Holzmann:2003}. Although the formal proof is rather involved, the main scaling result is intuitive, $n_0/n \propto \Delta n_c \propto a/\lo$ \cite{log}. From this result we immediately obtain $f_0 \propto (a/\lo)^4$, as illustrated in Fig.~\ref{fig:Cartoon}(b).
The presence of the harmonic trapping potential still affects the scaling of $f_0$ with $a/\lo$, but in this case the results for a harmonic and a uniform system are trivially related by the volume of the critical region, $\propto (a/\lo)^3$.

To experimentally measure $f_0$ we use an optically trapped cloud of $^{39}$K atoms in the $|F,m_F\rangle = |1,1\rangle$ hyperfine state, in which the strength of interactions can be tuned via a Feshbach resonance centred at 402.5 G \cite{Zaccanti:2009}. Our experimental system and the procedure for making precise and accurate measurements close to the critical point are described in detail in \cite{Campbell:2010, Tammuz:2011, Smith:2011}. Briefly, we prepare partially condensed clouds at various values of the scattering length $a$, and then let the number of atoms in the trap gradually decay through inelastic processes, while finite trap depth and sufficiently high rate of elastic collisions ensure that the sample remains in equilibrium at an approximately constant temperature. For the measurements presented here, $N \approx (4-5) \times 10^5$, the geometric mean of the trapping frequencies in our nearly isotropic trap is $\bar{\omega}/2\pi \approx 80\,$Hz, and $T \approx 250\,$nK, corresponding to $\lo \approx 10^4\,a_0$, where $a_0$ is the Bohr radius.

To discern condensed fractions as low as $\sim 0.1\,\%$ we turn off the interactions (switch $a$ close to zero) during TOF, thus minimizing the condensate expansion \cite{Smith:2011}. We numerically calculate $\Nmf$ using standard MF theory (see also \cite{details}) and measure the condensed atom number $N_0$ at the point where the total atom number is $N=\Nmf$.   To eliminate various sources of $a$-independent systematic errors (including absolute $N$ and $\bar{\omega}$ calibration) we perform ``reference" measurements in a weakly interacting gas with $a/\lo \approx 0.005$ \cite{Smith:2011}. At this reference point the expected value of $f_0$ is $< 10^{-5}$ (see below), and we neglect it in our analysis.

\begin{figure} [b]
\includegraphics[width=0.9\columnwidth]{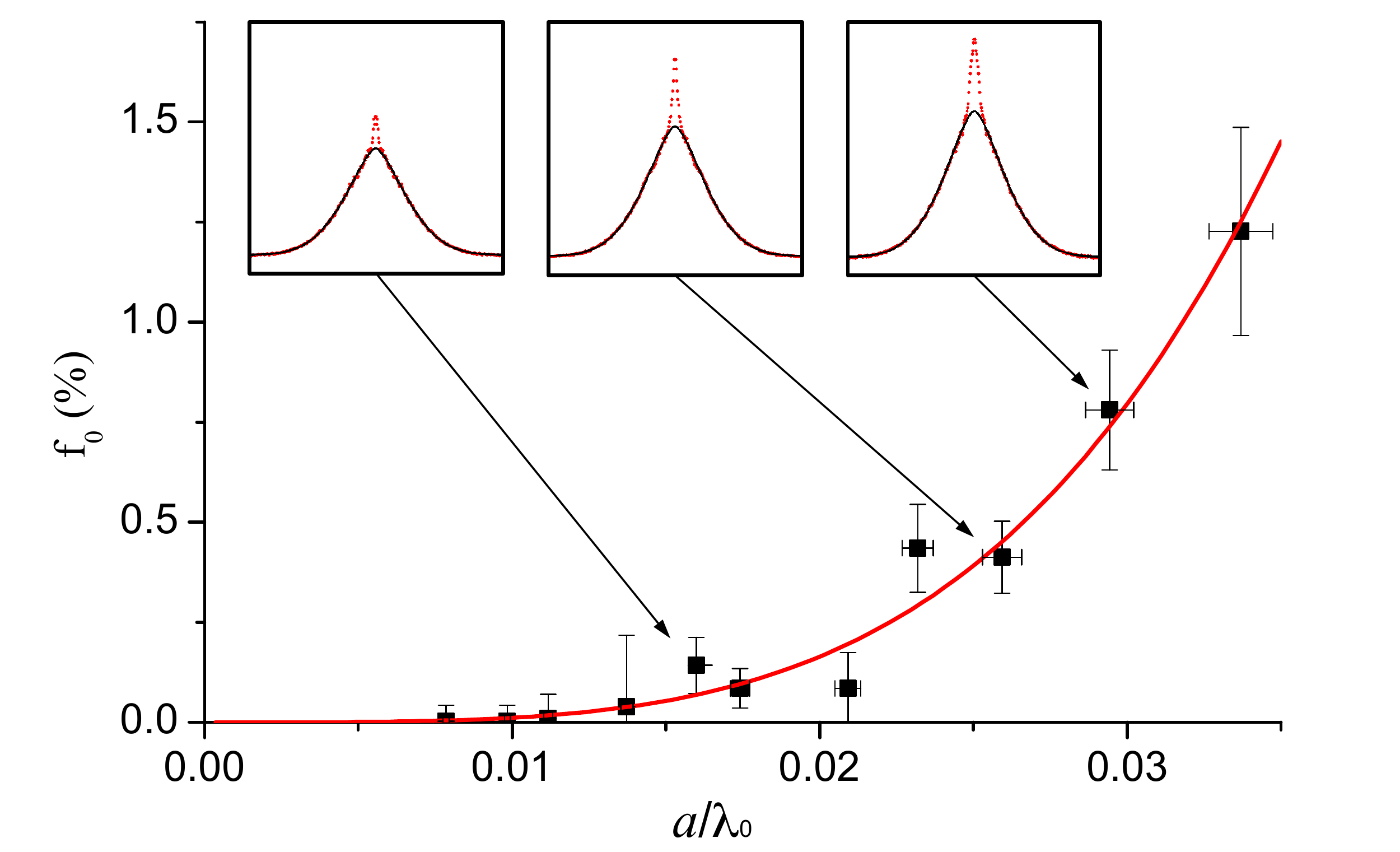}
\caption{(color online) Condensed fraction of an atomic gas induced by critical correlations. The condensed fraction $f_0$ is measured at the point where the total atom number is $N= \Nmf > N_c$. A fit to the data (solid line) with the function $f_0 \propto (a/\lo)^x$ gives an exponent $x=3.9 \pm 0.4$, in agreement with the predicted $x = 4$. Vertical error bars are statistical and horizontal error bars reflect the 0.1 G uncertainty in the position of the Feshbach resonance. The insets show representative  column density distributions after 19 ms TOF.}
\label{fig:quartic}
\end{figure}

Our experimental results are summarized in Fig.~\ref{fig:quartic}. Starting at zero for small $a$ (in agreement with MF theory), the condensed fraction $f_0$ grows to $\sim 1\%$ at $a \approx 350\,a_0$.
The use of a Feshbach resonance in principle allows us to increase $a$ further, but in the more strongly interacting gases the unfavorable ratio of the three-body loss rate to the two-body elastic collision rate precludes reliable equilibrium measurements \cite{Smith:2011}.

We fit our $f_0$ data with a function $(a/\lo)^x$ where $x$ is a free parameter. The fit yields $x = 3.9 \pm 0.4$, in agreement with the predicted $x=4$. This confirmation of the expected scaling of $f_0$ with $a/\lo$ is the first main result of this paper.

We now quantitatively relate our measurements to Monte-Carlo (MC) calculations for a uniform gas.
Following \cite{Prokofev:2004} we first define the reduced chemical potential
\begin{equation}
X = \frac{\mu-\mu_c}{32\pi^3(a/\lo)^2 k_B T} \; .
\end{equation}
Next, following \cite{Arnold:2001} we calculate $X_0$, the value of $X$ in the centre of the trap for $N = \Nmf$ (due to logarithmic corrections this is a slightly different condition from $\mu(0) = \mumf$, but this distinction is not experimentally observable).
We use the experimental value $b_2 = 42 \pm 2$, and $b_2^{\rm MF} = 11.7 \pm 0.1$ \cite{Smith:2011, details} to get
\begin{equation}
X_0  \approx  \frac{3\, \zeta(3)}{32\pi^3\zeta(2)} \left( b_2 -  b_2^{\rm MF} \right) = 0.067 \pm 0.005 \;  .
\end{equation}
For a uniform system the reduced condensate density $\tilde{f} (X)$, defined by $n_0 \lambda_0^3=16\pi^3 (a/\lambda_0)\tilde{f}(X)$, was tabulated in \cite{Prokofev:2004} using MC simulations.
Invoking LDA, for a harmonically trapped gas we get
\begin{equation}
\frac{N_0}{N_c^0} = \frac{\sqrt{2} (4\pi)^7}{4\zeta(3)}  \left(\frac{a}{\lo}\right)^4 \int_0^{X_0} \tilde{f}(X)\sqrt{X_0-X} \, d X \; .
\label{eq:MCLDA}
\end{equation}
Writing $(N_0 / N_c^0)^{1/4} = \alpha (a/\lo)$ and numerically evaluating the integral in Eq.~(\ref{eq:MCLDA}), using the results of \cite{Prokofev:2004}, we get the numerical coefficient
$\alpha_{\rm MC} = 10.4 \pm 0.4$.

\begin{figure} [b]
\includegraphics[width=0.9\columnwidth]{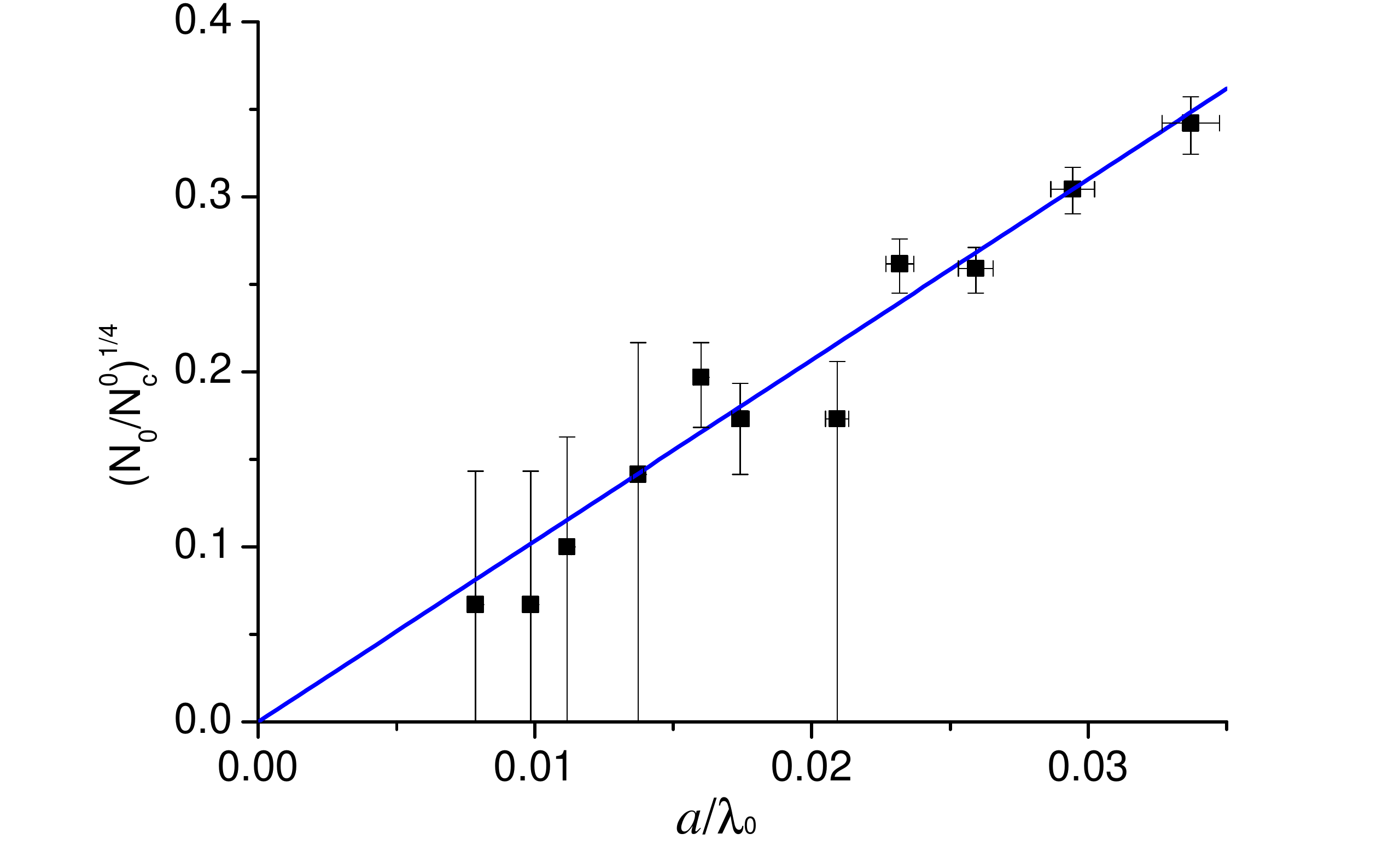}
\caption{(color online) Comparison with Monte-Carlo calculations for a uniform system. To quantitatively compare our data with the MC simulations we plot $(N_0 / N_c^0)^{1/4}$ versus $a/\lo$ (see text). A linear fit gives a gradient of $\alpha_{\rm exp} = 10.3\pm0.3$, in excellent agreement with the prediction $\alpha_{\rm MC} = 10.4\pm0.4$. The error bars are obtained using the limiting values from Fig. \ref{fig:quartic}; the points with large error bars do not significantly affect the fitted value of $\alpha_{\rm exp}$ but are clearly consistent with it.}
\label{fig:MC}
\end{figure}

In Eq.~(\ref{eq:MCLDA})  $N_0$ is calculated at $N=\Nmf$ but normalised to $N_c^0$. This expression therefore differs from $f_0$ by a factor $\Nmf/N_c^0$. This difference does not affect the leading $(a/\lo)^4$ term and is relevant only at the $(a/\lo)^5$ level. Nevertheless, for a direct quantitative comparison, in Fig.~\ref{fig:MC} we normalise the measured $N_0$ values to $N_c^0$, and assume the quartic dependence on $a/\lo$. The linear fit to  $(N_0 / N_c^0)^{1/4}$ yields the experimental value $\alpha_{\rm exp} = 10.3 \pm 0.3$, in excellent agreement with the Monte-Carlo result.

For another comparison, it is interesting to convert $X_0$ into $N_0$ using the standard Thomas-Fermi (TF) law. This MF law is valid well below $T_c$, where $N_0 \approx N$, but should not hold close to the critical point. For a given $X_0$, the TF law also predicts $N_0 \propto (a/\lo)^4$. However it corresponds to $\tilde{f}(X) = X$ and gives $\alpha_{\rm TF} = 8.2 \pm 0.4$. This result underestimates the condensed fraction $f_0$ by a factor $(\alpha_{\rm MC}/\alpha_{\rm TF})^4 \approx 2.6$, and we experimentally exclude it at about 4 sigma level.
This confirms that near $T_c$ mean-field theory fails on both sides of the critical point.

In conclusion, we have studied the condensed fraction of an atomic gas induced by inter-particle correlations at a point where no condensate is predicted by mean-field theory. Building on the recent observation of correlation effects on the condensation temperature of a trapped gas, this work makes a more direct connection with the critical behaviour in a homogeneous system. We experimentally confirm the predicted scaling $f_0 \propto (a/\lo)^4$, which highlights the conceptual difference between the interaction shifts of the critical density (characteristic of a uniform system) and the critical atom number in a harmonically confined cloud. Moreover, we demonstrate excellent quantitative agreement between our experiments and Monte-Carlo simulations for a homogeneous gas.
In a more general context, this provides an example of the potential of ultracold atomic gases for quantitative quantum simulation of intricate beyond-mean-field phenomena in uniform many-body systems.

We thank J. Dalibard for comments on the manuscript.
This work was supported by EPSRC (Grant No. EP/1010580/1). R.P.S. acknowledges support from the Newton Trust.

\end{document}